\documentclass[referee]{mn2e}

\input psfig.sty

\def\T1{\ {$T_1$}\ }
\def\MT1{\ {$M_{T_1}$}\ }
\def\ct1{\ {$(C-T_1)$}\ }
\def\CT10{\ {$(C-T_1)_0$}\ }

\def\VI0{\ {$(V-I)_0$}\ }

\def\2cd{\ {two-color diagram}\ }

\def\sf{\ {specific frequency}\ }
\def\ell{\ {elliptical}\ }

\def\gtsim{\ {\raise-0.5ex\hbox{$\buildrel>\over\sim$}}\ }
\def\ltsim{\ {\raise-0.5ex\hbox{$\buildrel<\over\sim$}}\ }

\begin{document}

\title[SMC candidate star clusters]{Washington photometry of candidate star clusters in the Small 
Magellanic Cloud}

\author[A.E. Piatti and E. Bica]{Andr\'es E. Piatti$^1$ and Eduardo Bica$^2$\\
$^1$Instituto de Astronom\'{\i}a y F\'{\i}sica del Espacio, CC 67, Suc. 
28, 1428, Ciudad de Buenos Aires, Argentina\\
$^2$ Universidade Federal do Rio Grande do Sul, Depto. de Astronom\'{\i}a, CP 15051, 
Porto Alegre, 91500-970, Brazil\\
}

\maketitle

\begin{abstract}

We present for the first time Washington $CT_1$ photometry for 11 unstudied or poorly studied candidate star clusters. 
The selected objects are of small angular size, contain a handful of stars, and are projected towards the innermost
regions of the Small Magellanic Cloud (SMC). The respective Colour-Magnitude Diagrams (CMDs) were cleaned from the
unavoidable star field contamination by taking advantage of a procedure which makes use of variable size Colour-Magnitude
Diagram cells.
This method has shown to be able to eliminate stochastic effects in the cluster CMDs
caused by the presence of isolated bright stars, as well as, to make a finer cleaning in the most populous CMD regions. 
Our results suggest that nearly 1/3 of the studied candidate 
star clusters would appear to be genuine physical systems. In this sense, the ages previously derived for some of them
mostly reflect those of the composite stellar populations of the SMC field. 
Finally, we used the spatial distribution in the SMC of possible non-clusters to statistically 
decontaminate that of the SMC cluster system. We found that there is no clear difference 
between both expected and observed cluster spatial distributions, although it would become more important 
at a 2$\sigma$ level between $a$ $\approx$ 0.3$\degr$ and 1.2$\degr$ (the semi-major axis of a ellipse parallel 
to the SMC bar and with $b/a$ = 1/2), if the asterisms were increased up to 20\%.

\end{abstract}


\section{Introduction}

The different catalogues of Small Magellanic Cloud (SMC) star clusters have been compiled on the basis of
star counts, either by visually inspecting photographic plates \cite[for example]{b75,h86,bs95} or by
automatic algorithmic searches \cite[for example]{petal99}. As far as we are aware, the most recent catalogue
which puts all the previous ones together is that of Bica et al. \shortcite[hereafter B08]{bietal08}. Although 
it is expected that most of the catalogued objects are indeed genuine physical systems, it 
was beyond the scope of Bica et al. (2008) to verify the physical nature of such faint objects.
The task of cleaning cluster catalogues from non physical systems or asterisms is far from being an exciting job.
Most of the astronomers desire to deal with prominent clusters. For this reason studies concluding about the asterism 
or overdensity nature of faint objects in the Clouds are rare or absent.
However, those works would be very important and are also required if any statistical analysis about the cluster 
formation and disruption rates, the cluster spatial, age and metallicity distributions, etc. is attempted.

As it is commonly accepted, an apparent 
concentration of stars in the sky does not necessarily 
lead to the conclusion that such concentration constitutes a physical system. The 
presence of such star concentration implies that we are dealing with a physical cluster
in the case of typical globular clusters or very populous open clusters. For most of the apparent star 
concentrations in the sky, however, it may be necessary to have supplementary information available 
about proper motions, radial velocities, spectral types and photometry to confirm their physical 
reality. The photometric data are often the only information at our disposal from which the 
existence of a star cluster may be inferred. 
Even though photometric data are indeed valuable, the steps to conclude on the physical nature of a star
aggregate from its Colour-Magnitude Diagram (CMD) might not be a straighforward task. This 
usually happens when dealing with small objects or sparse clusters  projected or inmersed in
crowded star fields. An example of such situations are those clusters located in the inner regions
of the SMC. In such cases, simple circular CMD extractions around the cluster centre could 
lead to a wrong
conclusion, since the CMDs are obviously composed of stars of different stellar 
populations \cite{p12}. Consequently, it is hardly possible to assess whether the bright and young
Main Sequence (MS) or the subgiant and red giant branches trace the fiducial cluster features.
Glatt et al. \shortcite[hereafter G10]{getal10} have studied 324 SMC clusters using data from
the Magellanic Cloud Photometric Surveys \cite{zetal02}. They show from isochrone fittings on to the CMDs 
that the studied  objects are clusters younger than 1 Gyr mostly distributed
in the main body of the galaxy, which is highly crowded. Although they mention that field contamination
is a severe effect in the extracted cluster CMDs and therefore influences the age estimates, no
decontamination from field CMDs were carried out. It would not be unexpected that some of the studied objects 
are not real star clusters, particularly those with very uncertain
age estimates ($\sigma$(age) $\ge$ 0.5 for log(age) $\la$ 9.0).  This possibility alerts us that the sole 
circular extraction of the observed CMDs of
clusters located in highly populated star fields is not enough neither for an accurate isochrone fitting 
to the cluster MSs nor for confirming their physical natures.

Different statistical procedures have been proposed with an acceptable success,
in order to avoid as much as possible the field contamination in the cluster CMDs. Chiosi et al. 
\shortcite[hereafter C06]{cetal06} studied 311 clusters in the central part of the SMC from OGLE data 
\cite{uetal98} and other own data. They used equivalent cluster areas of fields close to the clusters, but outside the
cluster radii, to build field CMDs. Subsequently, they divided the CMDs of both clusters and fields in boxes of 
size $\Delta$($V$)= 0.5 mag and $\Delta$($V-I$) = 0.2 mag, and subtracted for every field star in any box 
the closest cluster star in the respective box. The cluster ages derived from isochrone fittings on to
the cleaned CMDs for 136 clusters also included in the study of G10, resulted to be $\sim$ 0.2-0.3 in
log(age) younger than those by G10 ($\sigma$($\Delta$log(age)) = 0.13, age $\la$ 9.0). The main reason for this
systematic shift is probably the different metallicty of the isochrones involved. While C06 used the
isochrone set of Girardi et al. \shortcite[Z = 0.008]{getal02}, G10 fitted the cluster CMDs with
isochrones computed by Giradi et al. \shortcite[Z = 0.004]{getal95}. Note that for those clusters
with the most significant age deviation large age uncertainties are obtained in both C06 and G10 works.
Furthermore, bearing in mind 
the large age uncertainties quoted by C06 and G10 for some clusters, and that nobody has confirmed that
they are genuine physical systems, the doubt about their cluster reality might arise unavoidably.

In this paper we present an analysis of 11 candidate star clusters from new CCD
Washington $CT_1$ photometry, in combination with a computational tool for cleaning the
star field signature in the cluster CMDs. Our main aim is to be able to confirm the physical reality of the studied objects,
once their photometric data have been properly cleaned from field contamination. Indeed, the proposed
computational tool for estimating the probability of a star of being an intrinsic feature of the cluster field 
shows to be able to produce reliable CMDs revealing the genuine nature of the considered object. Note that the
studied objects were catalogued as clusters on the basis of star counts on less deep images than those using in this study.
We present the data set in Section 2, while we describe the data 
handling in Section 3. Section 4 deals, on the one hand, with available star field decontamination procedures  
 and, on the other hand, with the presently used method. 
In Section 5, our analysis shows that most of these stellar groups are likely genuine star clusters.
Finally, we summarize the main results in Section 6. 

\section{The data}

Piatti \shortcite{p12} performed the reduction, the stellar photometry and the photometric standardization
of Washington $CT_1$ images of eleven SMC fields obtained at the Cerro-Tololo Inter-American Observatory 
(CTIO) 4 m Blanco telescope with the Mosaic II camera attached (36$\arcmin$$\times$36$\arcmin$ field onto a
8K$\times$8K CCD detector array). The images are available at the National Optical Astronomy 
Observatory (NOAO) Science Data Management (SDM) Archives\footnote{http://www.noao.edu/sdm/archives.php.}.
When examining the CMDs of all the catalogued clusters observed in these images, we found 20 
intermediate-age or old clusters \cite{p11a,p11b},  69 moderately young or young clusters (forthcoming paper), 
41 clusters that fall between the image gaps and have not been measured, and 21 
candidate star clusters or asterisms. We are here focusing on to the last subgroup of objects. 
Table 1 lists 11 candidate star clusters along with the main astrometric, photometric and observational 
information. We recall that the $R$ filter used for imaging has a significantly higher throughput 
as compared with the standard Washington $T_1$ filter, so that $R$ magnitudes can be accurately transformed to yield 
$T_1$ \cite{g96}. We have also included the log(age) values obtained by C06 and G10, respectively, as well as their 
uncertainty
class, following the same notation as they used: class 1 indicates objects with age errors $\sigma$(log(age))
$<$ 0.3; class 2 indicates objects with age errors 0.3 $\le$ $\sigma$(log(age)) $<$ 0.5; and class 3 indicates
objects with age errors $\sigma$(log(age)) $\ge$ 0.5. The final information gathered for each candidate star
cluster of Table 1 consists of a running number per star, of the $x$ and $y$ coordinates, of the measured $T_1$ magnitudes
and $C-T_1$ colours, and of the observational errors $\sigma$($T_1$) and $\sigma$$(C-T_1)$. Table 2 gives
this information for B119. Only a portion of this table is shown here for guidance regarding its form and content. 
The whole content of Table 2, as well as the final information for the remaining candidate star clusters is available 
in the online version of the journal on Synergy, at http://www.blackwellpublishing.com/products/journals/suppmat/MNR/.
                                                   
As compared with the data set used by C06 and G10, the present $CT_1$ photometry looks deeper and more accurate.
The limiting magnitude of the photometry used by C06 is $V$ $\sim$ 21.5 mag. Most and least crowded fields are complete
down to $V$ = 20.5 mag at about 80\% and 89\% levels, respectively. The 50\% level is reached at about $V$ $\sim$ 
21.0 mag in the most crowded fields. Inside the radius of the densest clusters, the OGLE data are complete 
at 80\%, 70\% and 50\% levels at magnitudes $V$ = 19.0, 19.5, and 20.3, respectively. These $V$ values
correspond to MS turnoffs (TOs) of log(age) $\approx$ 8.3, 8.5, and 8.8, respectively, if the isochrone sets by
Girardi et al. \shortcite{getal95,getal02}, the $E(B-V)$ colour excesses and the ($m-M$)$_o$ distance modulus used
by C06 for the SMC are adopted, independently of the metallicty. On the other hand, the magnitude limit of the 
photometric survey used by G10 varies as a function of stellar crowding. Zaritsky et al. \shortcite{zetal02} found 
little  visible evidence for incompleteness for $V$ $<$ 20 mag, corresponding to a MS TO of log(age) $\approx$ 8.7, but 
the scan edges become visible when plotting the stellar surface density for stars with 20 $<$ $V$ $<$ 21. Furthermore, 
any statistical analysis of this catalogue fainter than $V$ $<$ 20 requires artificial star tests to determine 
incompleteness, which is becoming significant at these magnitudes. As for the present $CT_1$ data set, Piatti \shortcite{p12} 
performed different artificial star tests and found that in the most crowded fields the 100\% completeness level is 
reached at $T_1$ $\sim$ 22.0 mag, which corresponds to MS TOs of log(age) $\sim$ 9.6. MS TOs for stellar populations
of 1.4, 2.8, and 8.4 Gyrs are placed at $T_1$ $\approx$ 19.9 mag, 20.9 mag, and 21.9 mag, respectively. Thus,
we actually reach one magnitude below the MS TO of clusters younger than $\approx$ 1.5 Gyr, which is the age range
frequently found for clusters in the SMC inner disk and bar. We conclude that the photometric 
data sets used by C06 and G10 have been superseded by the present $CT_1$ photometric survey. 

\section{Data handling}

The centres and radii of the candidate star clusters are quantities of great importance, 
since it is possible to confuse a small cluster with a chance grouping of stars in these SMC crowded fields. With this
premise in mind, we performed a task as careful as possible of identification on to the images of the seleted objects 
and then, fixed their centres and radii. The whole catalogue by B08 was first mapped on to each image, so that each object 
in cluster pairs or groups of clusters could be identified without 
ambiguity. We carried out an interactive routine at the $C$ and $T_1$ images for each object separately, making use 
of the different tools provided by SAOImage DS9\footnote{SAOImage DS9 development has been made possible by funding 
from the Chandra X-ray Science Center (NAS8-03060) and the High Energy Astrophysics Science Archive Center 
(NCC5-568).} display programme. These display facilities allowed multiple frame buffers, region cursor manipulation,
many scale algorithms and colormaps, geometric markers, scaling, arbitrary zoom, rotation, pan, a variety of
coordinate systems, etc.

Since the appearance of the objects in the images looks like they may contain a handful of comparatively bright 
stars, or  one very bright star surrounded by faint stars, or only relatively faint stars, or a
combination of these three possibilities, the stellar density alone could not be a good indicator of the presence 
of a cluster. For this reason, we think that the above procedure of getting the centres and radii of the
studied objects is more meaningful from a physical point 
of view than performing statistical-gaussian fits on to their star distributions. Figs. 1 to 11 
show 5$\arcmin$$\times$5$\arcmin$ XDSS $B$ images of the selected objects (left panels) and a
captured enlargements of the $C$ images
centred on the studied candidate star clusters. Only Fig. 1 is shown here as an example; the remaining Figs. 
are available in the online version of the journal on Synergy, at 
http://www.blackwellpublishing.com/products/journals/suppmat/MNR/. We have drawn black circles with radii as
estimated by Bica et al. \shortcite{bietal08}. 
The figures present a close-up view of the studied objects. At first glance, some of these small 
candidate star clusters do not 
seem to be genuine physical systems. We note that the availability of a less deep photometric data set, like
those used by C06 and G10, could possibly mislead the cluster identification. The scale of our images is
0.274$\arcsec$ pix$^{-1}$, while those of the OGLE and the Magellanic Cloud Photometric Survey are 
0.42$\arcsec$ pix$^{-1}$  and 0.7$\arcsec$ pix$^{-1}$, respectively, so that crowding effects at the centre of the
objects is more important in their images.

\section{Star field decontamination methods}

\subsection{Background}

The issue about the decontamination of cluster fields from field stars has long been treated. Here we only mention
some techniques with the aim of summarizing the different approaches considered. 
Bonatto \& Bica \shortcite{bb07} developed a statistical method that basically involves: (i) dividing the full range 
of magnitude and colours of a given CMD into a 3D grid whose cubic cells have axes along the $J$, $J-H$ and $J-K_s$ 
directions, (ii) computing the expected number-density of field stars in each cell based on the number of comparison 
field stars with magnitude and colours compatible with those of the cell, and (iii) subtracting randomly the expected 
number of field stars from each cell. Bonatto \& Bica \shortcite{bb07} and subsequent studies estimate a radius where 
the subtracted CMD is statiscally more significant. With that well defined CMD, they build a magnitude-colour filter 
that is applied to the whole photometry, which produces the radial density distribution, if possible, well beyond the 
cluster limit. Although the method reapplies the cleaning 
procedure using different cell sizes in the CMDs, they are fixed each time, i.e., they do not vary across the CMDs. This
latter possibility could be very useful to eliminate field stars with small photometric errors located in CMD regions 
with a scarce number of stars (e.g. bright red giants, etc).

Pavani \& Bica \shortcite{pb07} and Pavani et al. \shortcite{pavanietal11}, based on Bonatto \& Bica´s \shortcite{bb07} 
method, further adapted the R$^2$ statistical test to uncover possible open cluster remnants, for which CMD field 
decontamination is priority. They tested whether the remnant candidate can be  reproduced  by equal area field 
fluctuations sorted out in the background area. This method is very suitable for sparsely populated objects.
Maia et al. \shortcite{metal10} also followed the precepts by Bonatto \& Bica \shortcite{bb07} 
to develop their own decontamination procedure. As above, they used cells in the CMDs with 
constant colour and magnitude bins for each size and positional configuration. They created an exclusion index by 
noting
how many times each star was removed from the sample, and then normalizing by the number of grid configurations.
On the other hand, they computed an average membership for each star. The decontamination procedure consists in 
applying complementarily both indicators. They imposed the restriction that the field CMD should have the same 
reddening as the cluster CMD.

With the aim of estimating a magnitude from which the characteristics of the observed field stars are indistinguishable 
from those of the clusters in terms of spatial density, magnitude, and colour distributions, Piatti et al.
\shortcite{petal10} applied a statistical method to filter the field stars from the CMDs and from the colour-colour 
diagrams. They divided the observed region into boxes with the same number of pixels per
side, and built for each of them the corresponding CMDs. At first glance, the method appears profitable when some 
differences can be seen between the various box-extracted CMDs, a fact which reveals a lack of homogeneity in the 
spatial distribution of the brightness and colours of the stars. The method used consists in alternatively adopting 
any of the box-extracted "reference" CMDs 
to statistically filter the remaining ones. This filtering task is repeated using each of the box-extracted CMDs as a 
reference CMD. In the end of this process, each box-extracted CMD was individually filtered, using each time a 
different reference CMD. The filtering was performed in such a way that the stars in different magnitude-colour 
bins for each reference CMD are subtracted from the remaining CMDs by removing those stars closer in magnitude and 
colour to the ones of the reference CMD. 
When comparing the various filtered CMDs corresponding to a given box with the observed one, the residuals from 
box-to-box variations and the fiducial CMD features of that box are found. This is due to the
fact that a star that has magnitude and colours within the typical values found in the reference field CMDs is
in most cases eliminated. Thus, the fewer times a star is removed in a given box, the larger its
probability of representing a fiducial feature in that box. Piatti et al. \shortcite{petal10} adopted any star that 
was removed fewer than 20\% of the times as a probable fiducial feature star. They found in this way a magnitude from 
which it is hardly possible distinguishing cluster stars from field stars.

\subsection{An alternative cleaning procedure}

From our experience in cleaning the field star contamination in the cluster CMDs, we have identified some situations
which still need our attention. It frequently happens that some parts of the CMDs are more populated than others, so 
that fixing the size of the cells in the CMDs becomes a difficult task. Small cells do not usually carry out a 
satisfactory job in CMD regions with a scarce number of fields stars, while big cells fail in populous CMD regions. 
Thus, relatively
bright field red giants with small photometric errors could not be subtracted and, consequently, the cluster
CMD could show spurious red giant features.  A compromise 
between minimizing the residuals left after the subtractions of field stars from the cluster CMDs and maximizing the 
cleaning of field stars is always desiderable. 
This fact has 
led previous procedures to try with different cell sizes 
as a general budget which, in turn, depends on the sizes of the spatial regions used to build the extracted CMDs.
Sometimes the radial density profile of a star cluster is taken into account to assign statistical membership 
probabilities. This constrains the star field cleaning procedure to those clusters for which it is possible to 
satisfactorily trace their stellar radial distributions. Unfortunately, this is not the case for objects with
a handful of stars or of small angular dimensions. 

We have designed an alternative procedure which makes use of variable cells in the CMDs. Magnitude and colour
cell sizes are varied separately. The cells are adjusted in such a way that they result bigger in
CMD regions with a scarce number of stars, and viceversa. This way, we pursue to map the field CMD as closely as
possible on to the cluster CMD. The method does not need to know whether a star is placed close to the cluster centre
nor the cluster radial density profile to infer a membership probability. However, it takes into account the star field
density, since the more populous a star field the larger the number of stars subtracted from the cluster CMD. This is 
accomplished by eliminating one star -the closest one in terms of magnitude and colour- in the cluster CMD for each 
star identified in the field CMD. As a result, the intrinsic spatial star distribution is uncovered within the object 
region.

Once the field CMD is adopted, the method defines a free path for each star as the distance to the closest
star in the field CMD. Magnitude and colour directions are separately considered, so that ($\Delta$(colour))$^2$ +
($\Delta$(magnitude))$^2$  = (free path)$^2$, where $\Delta$(colour) and $\Delta$(magnitude) are the distances 
from the considered star to the closest one in abscissa and ordinate in the field CMD. The method has shown to be 
able to eliminate stochastic effects in the cluster CMDs
caused by the presence of isolated bright stars, as well as, to make a finer cleaning in the most populous CMD
regions. In order to prevent from large non-meaningful free paths, the method 
imposes a reasonably large free path limit. The free path of a field star accounts for a zone of influence 
(rectangle) of that star in the CMD, in the sense that only the closest star inside that area in the cluster CMD is 
eliminated. 

For our list of candidate star clusters we cleaned circular regions centred on the objects with radii twice as big as those 
estimated by Bica et al. \shortcite{bietal08} for the objects, i.e, we cleaned areas four times bigger than those of the 
circles of Figs.
1 to 12. As for the reference star fields, we used four different regions per object with the same area as for
the candidate clusters. These circular field areas were placed towards the north, the east, the west and the south of 
the candidate cluster at a distance of four times the radii of Figs. 1 to 12. This was done in order to take into account 
variations in the spatial density, magnitudes, and colours of field stars. Note that the candidate star clusters are of
small angular sizes, typically $\sim$ 0.1$\arcmin$-0.3$\arcmin$. We previously performed a large number of tests
by using as object and field CMDs a unique selected field, or selected two different field CMDs, one of them
acting as cluster CMDs. The results of our various experiments showed that residuals are not left or they are 
minimal. Anyway, we do not rule out more noticeable residuals due to the non uniformity of the field in terms 
of spatial density, colour distribution and luminosity function. However, whenever an
excess of stars remains in the cleaned CMDs, we assume that we are dealing with an enhancement of stars 
caused by the presence of a possible star cluster or by a stellar fluctuation in the SMC field or by a chance grouping 
of stars. 

The method was run four times for each studied object, each time using a different reference field area.
Thus, we obtained four different cleaned CMDs per object. When comparing those CMDs, one may find 
stars that have kept unsubtracted in most of the times, while other stars were subtracted in most of the program executions.
The different number of times that a star keeps unsubtracted can then be converted in a measure of the probability of being a
fiducial feature of the candidate cluster field. Thus, we are able to distinguish stellar populations projected on to
the cluster fields that have a probability P $<$ 25\% of being a genuine candidate cluster population, i.e., a typical 
foreground population; stars that could indistinguishably belong to the star field or to the studied object (P = 50\%);
and stars that are predominatly found towards the candidate cluster field (P $>$ 75\%) rather than in the star field
population.

\subsection{Results}

Figs. 12 to 22 
show in the top-right panels the extracted field CMDs.  Only Fig. 12 is shown here as an example; the remaining Figs. 
are available in the online version of the journal on Synergy, at 
http://www.blackwellpublishing.com/products/journals/suppmat/MNR/. Instead of plotting the position of each
field star, the graphics present rectangles -which come from the definition of a free path
for each star- centred on those positions, with the aim of illustrating how the technique works. As can be seen, this 
technique allows using 
smaller and larger rectangles according to the colour and magnitude frequencies in the CMDs. We think that 
the flexibility provided by variable 
cells can reproduce more tightly the field CMD features on to the candidate cluster CMDs than those of fixed sizes, and thus
allowing improvements in the field star cleaning process. The top-left panels depicts schematic finding charts including all
the measured stars drawn using open circles, whose sizes are proportional to the $T_1$ brighness of the stars. The stars
with filled black, dark-gray and clear-gray circles have probabilities of being an intrinsic candidate cluster feature higher than 75\%, equal to 50\% and lower than 25\%, 
respectively. The circle represents the adopted radius which, in some cases, is
slightly shifted or resized as compared to that given in Bica et al. \shortcite{bietal08}. This was done in order to encompass stars which better define the field 
CMD features.
The bottom panels of the figures depict stars located within the overplotted circles for three different
probabilities regimes. The composite CMD from these three CMDs results in the observed candidate cluster CMD.

By comparing the three bottom CMDs we can find out whether a sequence or group of stars comes from the superposition of
different field populations or is part of the candidate cluster population. Some hints at this stage are easy to recognize.
We have overplotted on to the bottom panels of Figs. 12 to 22 the isochrones 
adopted by C06 and G10, respectively drawn with solid and dashed lines. We have used their metallicities, 
$E(B-V)$ colour excesses and SMC distance modulus. As for comparison purposes, we have also included 
the Zero Age Main Sequences (ZAMSs, $Z$ = 0.008, Piatti 2011b) with solid lines in all the CMDs, using a SMC distance 
modulus of $m-M$ = 18.9$\pm$0.1 and the $E(B-V)$ colour excesses taken from the Burstein \& Heiles 
\shortcite[hereafter BH]{bh82} 
extinction map. Whenever possible, we also estimate the cluster ages by fitting the isochrones of Girardi et al. 
\shortcite{getal02}, which are shown with dotted lines. We provide below with details about the studied candidate star 
clusters :

\noindent {\it B119:} The P($<$25\%) CMD shows the oldest field population that we find along the line-of-sight towards 
the object. According to the $\delta$($T_1$) index -which measures the difference in $T_1$ magnitudes between the Red Clump
(RC) and the MS TO- and equation (4) from Geisler et al. \shortcite{getal97}, its age is $\sim$ 2.5-3.0 Gyr. 
Note that some hints of such an evolved population is also seen in the P(=50\%) and P($>$75\%) CMDs. 
These latter CMDs also show a relatively straight MS which does not resemble that of a moderately young cluster.  
Since multiple populations appear in the three CMDs, we do not classify this object as a possible cluster.


\noindent {\it BS20:} RC stars seems to appear in the three bottom CMDs. They at $\approx$ 18.0 $\pm$ 0.5 mag. 
We discard the possibility of an intermediate-age cluster ($t$ $>$ 1 Gyr), since we would had expected to see its RC at a 
lower $T_1$ magnitude. For instance, RC stars older than 1.0 Gyr are placed at $T_1$ $\sim$ 18.8-19.1 
mag, depending on the stars' positions due to the galaxy depth \cite{cetal01}. In this case, the cluster would be far away 
from the SMC limits, between the SMC and our Galaxy. In addition, if we adopt $T_1$ $\sim$ 18.2 mag for the cluster 
MS TO, then the $\delta$($T_1$) index turns out ot be $\ga$ 2.3, which implies an age older than 3.5 Gyr. 
It is hardly possible that BS20 is an old SMC cluster \cite{p11b}. Another possibility is that the object be
a moderately young cluster, if we assume that the RC stars belong to the object. In such a case, an isochrones of log($t$) 
= 8.55 ($E(B-V)$ = 0.035, $Z$ = 0.008) would fit better the RC than C06 and G10 did. 
We might be dealing with a chance grouping of stars composed by a handful 
of faint MS and a few RC field stars.

\noindent {\it BS25:}  The P($>$75\%) and P($<$25\%) CMDs are almost identical, with 
the only difference being the few stars brighter than $T_1$ $<$ 20 mag and 
bluer than $C-T_1$ $<$ 0.2 mag. Old field MS stars
($T_1$ $>$ 21 mag) contaminate the entire candidate cluster field, as judged by the presence of them in the three bottom CMDs. Note also that a possible subgiant branch at $T_1$ = 20
is also visible. On the other hand, the P(=50\%) CMD shows a straight MS which does
not match the lower envelope of any young cluster MS. Based on the present observational limitations, we are inclined
to conclude that we are dealing with a possible overdensity of faint stars,
with superimposed few brighter sources.

\noindent {\it BS35:} As far as the stellar populations in the three bottom CMDs are compared, we note that
the star fields far outside from and inside the object circle are distinguishable. The cluster appears to be
 slightly younger than C06 and G10 estimated. Indeed, the isochrone which best reproduces the P($>$75\%) CMD's 
features is that
of log ($t$) =  8.70$\pm$0.05 ($E(B-V)$ = 0.014, $Z$ = 0.008).  Note that the MS TOs of the isochrones
adopted by C06 and G10 are redder than that for the object.

\noindent {\it BS251:}  At first glance, it would appear that the small angular size and low density candidate cluster is projected on to a much older and 
denser star field, in very good agreement with the age obtained by C06. However, the
space distribution of stars with P($>$75\%) and P(=50\%) is not as concentrated as
that for the P($<$25\%) stars. Based on this feature, we conclude that they are rather a star fluctuation in the field.

\noindent {\it BS265:} The cluster arises clearer in the  P($>$75\%) CMD where we fitted an isochrone for log($t$) = 8.8
$\pm$ 0.1, as G10 did, but using a different metallicity value ($Z$ = 0.008). The $E(B-V)$ obtained from BH is 0.068 mag.

\noindent {\it H86-70:} This is one of the cases where the cluster appears projected on to a star field
blurring its CMD's fiducial features. However, the cleaning procedure allows us to
disentagle them. The bottom-left panel shows a mixture of younger and older field MS stars and RC
field stars, while the bottom-right panel depicts a clear relatively tight cluster MS and, presumibly,
some cluster RC stars. Since field and cluster stars appear to have similar ages, C06 and G10 estimated
a cluster age in excellent agreement with our present  estimate (log($t$) = 8.8$\pm$0.1, $E(B-V)$ = 0.035, 
$Z$ = 0.008).

\noindent {\it H86-78s:}  This is a small angular size candidate cluster composed by a few bright stars, since
they only appear in the P($>$75\%) CMD; the star field is seen in the three
bottom CMDs and is clearly older. 
Since the apparent two lower cluster MS stars in the P($>$75\%) CMD are likely field stars, we
conclude that this object is a chance grouping of stars.

\noindent {\it H86-86:} Despite the visible excess of subgiant and red giant stars in the three bottom
panels -which resembles older field SMC populations-, the positions of the MS stars are reproduced by an 
isochrone of log(age) = 8.1 (G10), satisfactorily. Note, however, that such MS stars are not mainly seen
in the  P($>$75\%) CMD. We are possibly dealing with a chance grouping of stars. Therefore, the
isochrone fit performed by G10 would represent the age of the younger stellar population seen towards this
line-of-sight. 

\noindent {\it H86-88:} The catalogued object does not result to be a clear star cluster but rather 
a chance grouping of a few bright and a handful of low MS field stars. The composite field CMD features
led C06 and G10 to fit very different isochrones to the observed CMD.

\noindent {\it H86-197:} We confirm the physical reality of this intermediate-age cluster located in the
outer SMC disk. This is the first time that an age estiamted in provided for the object. We find that the
isochrones which best resembles the cluster CMD feature is that for log($t$) = 9.1$\pm$0.1 ($E(B-V)$ = 
0.024, $Z$ = 0.008).


Our results suggest that  nearly 1/3 of the studied candidate star clusters would appear to 
be genuine physical systems.  In addition, we find in most of the studied objects that there are more stars
with P($>$75\%) outside the cluster circle than inside.
The photometric errors of stars with probabilities of being a fiducial feature of the candidate cluster field
higher than 75\% are at most twice as big as the plotted filled circles in $C-T_1$ colour and smaller than the 
filled circles in $T_1$ magnitude. This means that the colour dispersion observed in some cleaned CMDs comes from an 
intrinsic dispersion in the colours of the involved stars. On the other hand, our study shows that the ages derived by 
C06 and G10 can reflect those of the composite stellar populations of the SMC field. As far as we are aware,
this is the first time that evidence is presented showing that some SMC candidate star clusters are not possible
genuine physical systems. We have additionally studied other ten candidate star clusters (B110, B112,
BS38, H86-78n, H86-91, H86-95, H86-96, H86-98, H86-103, H86-121), following the same precepts
as described for objects of Table 1, and also found that they are possible non-clusters. However, we cannot  
draw final conclusions on their status  since the images from which the photometry was obtained
contain some saturated stars. 

The resulting possible non-cluster sample represents nearly 10\% of the 152 objects
placed throughout the 11 observed SMC fields (see Sect. 2). This does not seem to be a significative percentage of the
catalogued clusters. If we assume a similar percentage for the whole catalogued SMC clusters \cite{p11b}, 
we would expect that some $\sim$ 60 objects were not real stellar aggregates. However, since these objects have
their own spatial distribution, it would be interesting to recover the expected
spatial distribution of genuine star clusters as a perspective exercise. Thus, the spatial distribution of possible 
non-clusters is used to correct that of the clusters in the observed 11 SMC fields
 in order to obtain an intrinsic spatial distibution. Then, by assuming that the area covered by these 11 
Mosaic II fields represents an unbiased subsample of the SMC as a whole, the expected spatial
distribution of the SMC cluster system is obtained, once it is statistically decontaminated by the spatial distribution of
the possible non-clusters. 

Viewing the SMC as a triaxial galaxy with the declination, right ascension and line-of-sight as the three axes, 
Crowl et al. \shortcite{cetal01} found axial ratios of approximately 1:2:4. Based on this
result, and with the purpose of describing the spatial distribution of
the clusters, we decided to use an elliptical framework instead of
a spherical one in order to reflect more meaningfully the flattening of the galaxy \cite{p11b}. Then, here, $a$ is the 
semimajor 
axis - parallel to the SMC main body - of an ellipse centred at RA = 00h 52m 45s, Dec. = −72$\degr$ 49$\arcmin$ 43$\arcsec$
(J2000) \cite{cetal01} and with a $b/a$ ratio of 1/2. Thus, we assume that the cluster spatial distribution correlates much 
better with a pseudo-elliptical (projected) distance measured from the galaxy centre than with the radial distance,
or distances defined along the right ascension or declination axes. For the subsequent analysis, we adopt the semimajor axis 
$a$ as the representative spatial variable to trace the behaviour of the cluster spatial distribution throughout the galaxy. 

We computed for each cluster in B08 the value of $a$ that an ellipse would
have if it were centred on the SMC centre, had a $b/a$ ratio of 1/2 and
one point of its trajectory coincided with the cluster position. We assume for the SMC the limits quoted by B08.
We counted the number of clusters in elliptical rings from $a$ = 0$\degr$ up to 8$\degr$. The size of the rings was
varied from 0.05$\degr$ up to 0.5$\degr$, so that we built a total of 10 cluster spatial distributions. 
Binning data where the distribution changes steeply frequently 
have biases in the measured distribution. In this case it  tends to 
pull the central spatial distribution remarkably high, since more clusters migrate into a given bin
from an adjacent bin where there are more clusters, than from the other 
adjacent bin where there are fewer clusters. The errors increase as one goes outside-in towards the SMC centre. 
We averaged the 10 resulting spatial distributions in order to mitigate the influence
of any chosen bin size. We repeated the same procedure to build the spatial distribution of non-possible
clusters as well as that for the observed ones in the 11 studied SMC fields. Then, we calculated the
expected number of clusters in the whole galaxy, corrected for possible asterisms in the B08's catalog according to
the expression:

\begin{equation}
N_{total} = (N_{obs} - N_{nocls})*N_{cat}/N_{obs}
\end{equation}

\noindent where $N_{nocls}$, $N_{obs}$, and $N_{cat}$ represent the spatial distributions of possible non-clusters, of
observed clusters in the 11 SMC fields, and of catalogued clusters in B08, respectively.

The inner panel of  Fig. 23 depicts the B08 cluster distribution (dots) and two ellipses of $a$ = 0.5$\degr$ and
1.0$\degr$. The large boxes represent the studied Mosaic II fields. One of them, centred on the cluster AM\,3, falls
beyond the figure limits at (relative R.A.$\times$cos(Dec),relative Dec.) $\approx$ (-4.7$\degr$,0.0$\sim$).
The larger panel shows both cluster spatial distributions, the one directly obtained from B08
(dotted histogram) and that decontaminated from possible non-clusters (solid histogram). The errorbars represent
a measure of the effect caused by choosing different bin sizes when building the cluster spatial distribution. 
As can be seen, there is no clear difference between both expected and observed cluster spatial distributions
within 1$\sigma$, although it would become to be important (a histogram difference bigger than 2$\sigma$)
between $a$ $\approx$ 0.3$\degr$ and 1.2$\degr$, if the asterisms were increased up to 20\%.

\section{Summary}

To date, the catalogue by Bica et al. \shortcite{bietal08} has been the most complete compilation of star clusters in the 
SMC. Most of these objects have not been studied yet. Here, we present for the first time Washington $CT_1$ photometry
for 11 unstudied or poorly studied candidate star clusters. As compared with the data sets from previous photometric 
surveys, the present $CT_1$ photometry turns out to be deeper and more accurate.  In general the selected objects appear to 
be of small angular size and contain a handful of stars. They are projected towards the most crowded
star field regions in the SMC, at distances shorter than $\sim$ 1$\degr$ from its centre.

We have designed a procedure for cleaning the cluster CMDs from the unavoidable star field contamination
which makes use of variable cells in the CMDs. The cells are adjusted in such a way that they result bigger in
CMD regions with a scarce number of field stars, and viceversa. This way, we reproduce the field CMD as closely as
possible on to the cluster CMD. The method does not need to know whether a star is placed close to the cluster centre
nor the cluster radial density profile to infer a membership probability. However, it takes into account the star field
density, since the more populous a star field the larger the number of stars subtracted from the cluster CMD. 
As a result, the intrinsic spatial star distribution is uncovered within the object region.
Once the field CMD is adopted, the method defines a free path for each field star as the distance to the closest
star in the field CMD. The method has shown to be able to eliminate stochastic effects in the cluster CMDs
caused by the presence of isolated bright stars, as well as, to make a finer cleaning in the most populous CMD regions. 

When applying the cleaning procedure to the CMDs of the 11 selected candidate star clusters, we found that
  nearly 1/3 
of them
would appear to be genuine physical systems. We estimated their ages  from the
matching of the isochrone which best reproduces the CMD cluster features. In this sense, the ages previously derived 
for some of them mostly reflect those of the composite stellar populations of the SMC field. 
The present analysis tools applied to  faint poorly populated clusters or candidates in the Magellanic Clouds
points  to the need of  better scale deep observations with e.g. the 8m class telescopes.

Finally, we used the spatial distribution in the SMC of possible non-clusters to statiscically 
decontaminate that of the SMC cluster system. By assuming that the area covered by 11 studied fields 
(36$\arcmin$$\times$36$\arcmin$ each) represents an unbiased subsample of the SMC as a whole and by using an elliptical
framework centred on the SMC centre ($b/a$ = 1/2), we found that there is no  significant difference between the
expected and the observed cluster spatial distributions. However, a difference at a 2$\sigma$ level would become 
visible between $a$ $\approx$ 0.3$\degr$ and 1.2$\degr$, if we doubled the amount of possible non-clusters.

\section*{Acknowledgements}

 We greatly appreciate the comments and suggestions raised by the
reviewer which helped us to improve the manuscript.
This research draws upon data distributed by the 
NOAO Science Archive, and has made use of SAOImage DS9, developed 
by Smithsonian Astrophysical Observatory. NOAO is operated by the Association of Universities for 
Research in Astronomy (AURA), Inc. under a cooperative agreement with the National 
Science Foundation. This work was partially supported by the Argentinian institutions CONICET and
Agencia Nacional de Promoci\'on Cient\'{\i}fica y Tecnol\'ogica (ANPCyT), and the Brazilian Agency CNPq.

\clearpage

\begin{table}
\caption{Candidate star clusters in the SMC.}
\begin{tabular}{@{}lcccccccccccc}\hline

Name & $\alpha_{\rm 2000}$ & $\delta_{\rm 2000}$ & {\it l}  & b         & date & exposure & airmass & seeing  & 
\multicolumn{4}{c}{log(age)}\\
     & (h m s)  & ($\degr$ $\arcmin$ $\arcsec$) & ($\degr$) & ($\degr$) &      & $C$ $R$ (sec) &  
$C$ $R$ & $C$ $R$ ($\arcsec$) & \multicolumn{2}{c}{C06} &\multicolumn{2}{c}{G10} \\\hline

B119     & 01 04 19 & -73 09 54 & 301.63 & -43.93 & 2008 Dec 19 & 1500 300 & 1.416 1.403 & 1.2 0.9 & --- &-- & --- &--\\
BS20     & 00 43 38 & -72 58 48 & 303.72 & -44.13 & 2008 Dec 20 & 1500 300 & 1.533 1.470 & 1.2 1.1 & 8.8 & 1 & 8.9 & 1\\
BS25     & 00 44 46 & -72 53 53 & 303.61 & -44.22 & 2008 Dec 20 & 1500 300 & 1.533 1.470 & 1.2 1.1 & --- &-- & --- &--\\
BS35     & 00 47 50 & -73 28 42 & 303.28 & -43.64 & 2008 Dec 20 & 1500 300 & 1.533 1.470 & 1.2 1.1 & 8.5 & 1 & 8.55& 1\\
BS251    & 00 51 26 & -73 17 00 & 302.93 & -43.84 & 2008 Dec 20 & 1500 300 & 1.533 1.470 & 1.2 1.1 & 8.4 & 2 & --- &--\\
BS265    & 00 57 22 & -71 53 29 & 302.27 & -45.22 & 2008 Dec 18 & 1500 300 & 1.595 1.559 & 1.3 1.1 & --- &-- & 8.8 & 2\\
H86-70   & 00 43 44 & -72 58 36 & 303.71 & -44.13 & 2008 Dec 20 & 1500 300 & 1.533 1.470 & 1.2 1.1 & 8.8 & 1 & 8.9 & 1\\
H86-78s  & 00 46 12 & -73 23 39 & 303.44 & -43.72 & 2008 Dec 20 & 1500 300 & 1.533 1.470 & 1.2 1.1 & --- &-- & 8.0 & 1\\
H86-86   & 00 47 01 & -73 23 35 & 303.36 & -43.73 & 2008 Dec 20 & 1500 300 & 1.533 1.470 & 1.2 1.1 & --- &-- & 8.1 & 1\\
H86-88   & 00 46 58 & -73 20 09 & 303.37 & -43.78 & 2008 Dec 20 & 1500 300 & 1.533 1.470 & 1.2 1.1 & 8.8 & 3 & 8.5 & 1\\
H86-197  & 01 15 30 & -71 10 44 & 300.15 & -45.81 & 2008 Dec 20 & 1500 300 & 1.533 1.470 & 1.2 1.1 & --- &-- & --- &--\\

\hline
\end{tabular}
\end{table}

\begin{table}
\caption{CCD $CT_1$ data of stars in the field of B119.}
\begin{tabular}{@{}lcccccc}\hline
Star & $x$  & $y$ & $T_1$ & $\sigma$$(T_1)$ & $C-T_1$ & $\sigma$$(C-T_1)$ \\
     & (pixel) & (pixel) & (mag) & (mag) & (mag) & (mag)  \\\hline
-    &   -     &   -     &  -    &  -    &    -   &   -      \\
     10 &3841.42 &7168.93  &20.938   &0.021   &0.158   &0.025\\
     11 &3841.61 &7194.83  &21.242   &0.043   &0.565   &0.046\\
     12 &3841.88 &7187.26  &20.130   &0.016   &0.384   &0.026\\
-    &   -     &   -     &  -    &  -    &   -   &   -      \\
\hline
\end{tabular}
\end{table}

\begin{figure}
\centerline{\psfig{figure=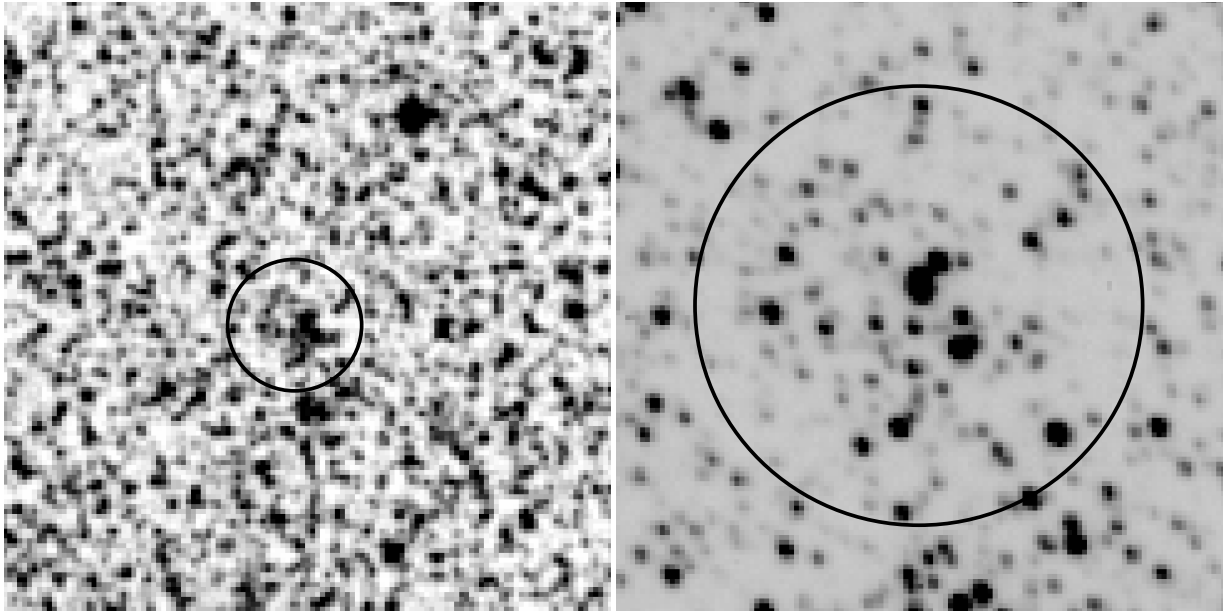,width=162mm}}
\caption{5$\arcmin$$\times$5$\arcmin$ XDSS $B$ image centred on B119 (left panel). North is up and east is to 
the left. The superimposed circle corresponds to the radius adopted by Bica et al. (2008). An enlargement of 
our $C$ image centred on the object is shown in the right panel.}
\end{figure}











\setcounter{figure}{11}
\begin{figure}
\centerline{\psfig{figure=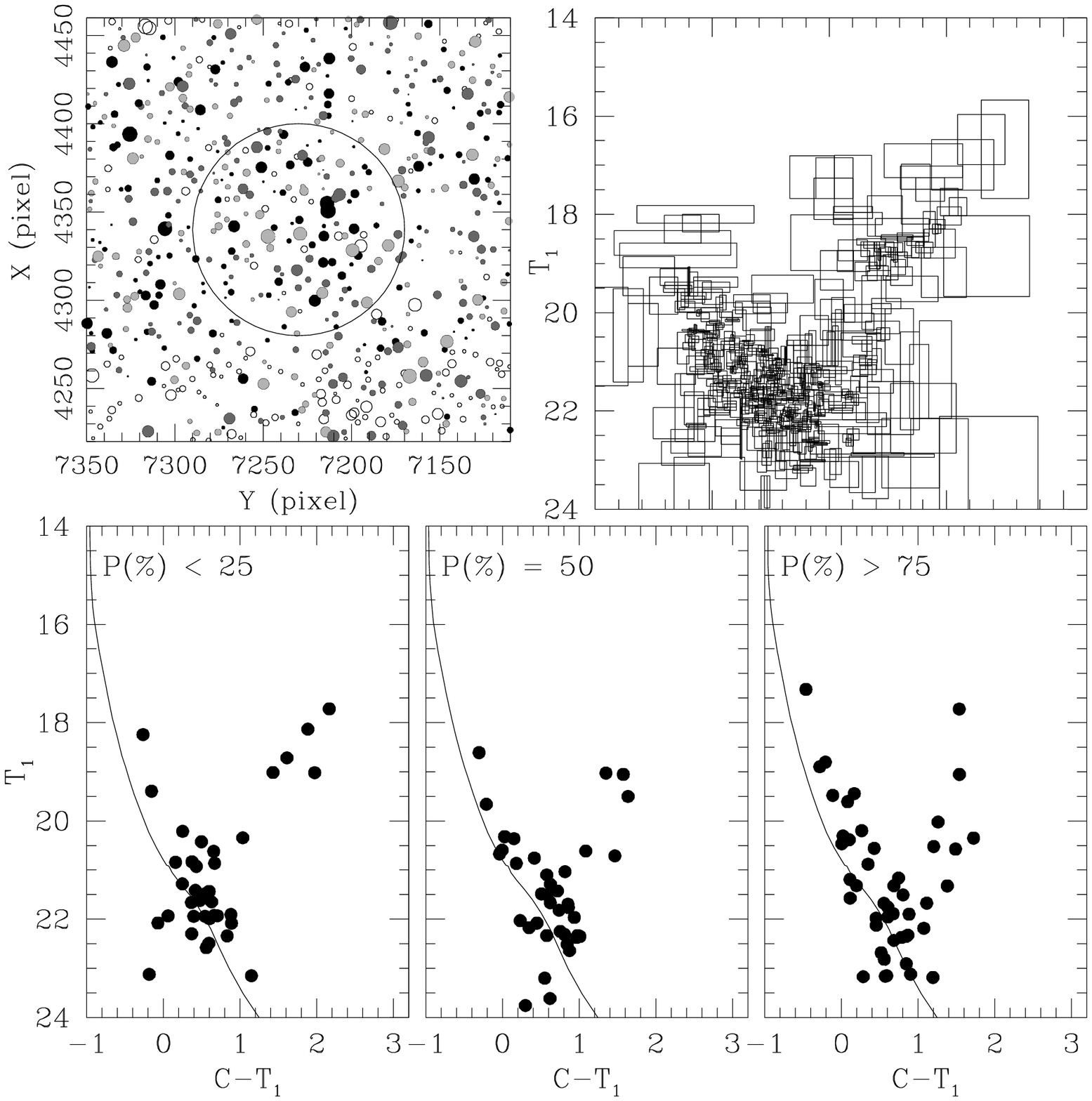,width=162mm}}
\caption{Schematic finding chart of the stars measured in the field of B119 (upper left), 
with a circle corresponding to the  adopted radius. North is up and east is to the left. 
The size of the symbols are proportional to the $T_1$ mag. Filled circles correspond to stars with a 
probability of being a feature of the cluster field higher than 75\%. The equal cluster area field CMD is
shown at the top-right panel, wherein the stars are placed at the centres of their respective free path rectangles.
For stars inside the candidate cluster radius, three different CMDs are shown, distinguishing those
stars that have chances of being a candidate cluster field feature  $<$ 25\%, equal to 
50\%, and $>$ 25\%,
respectively. The ZAMS (solid line) and the isochrones from Girardi et al. (2002) for the ages derived 
by C06 (solid line) and G10 (dashed line) are superimposed (see details in Section 4).}
\end{figure}

\setcounter{figure}{22}
\begin{figure}
\centerline{\psfig{figure=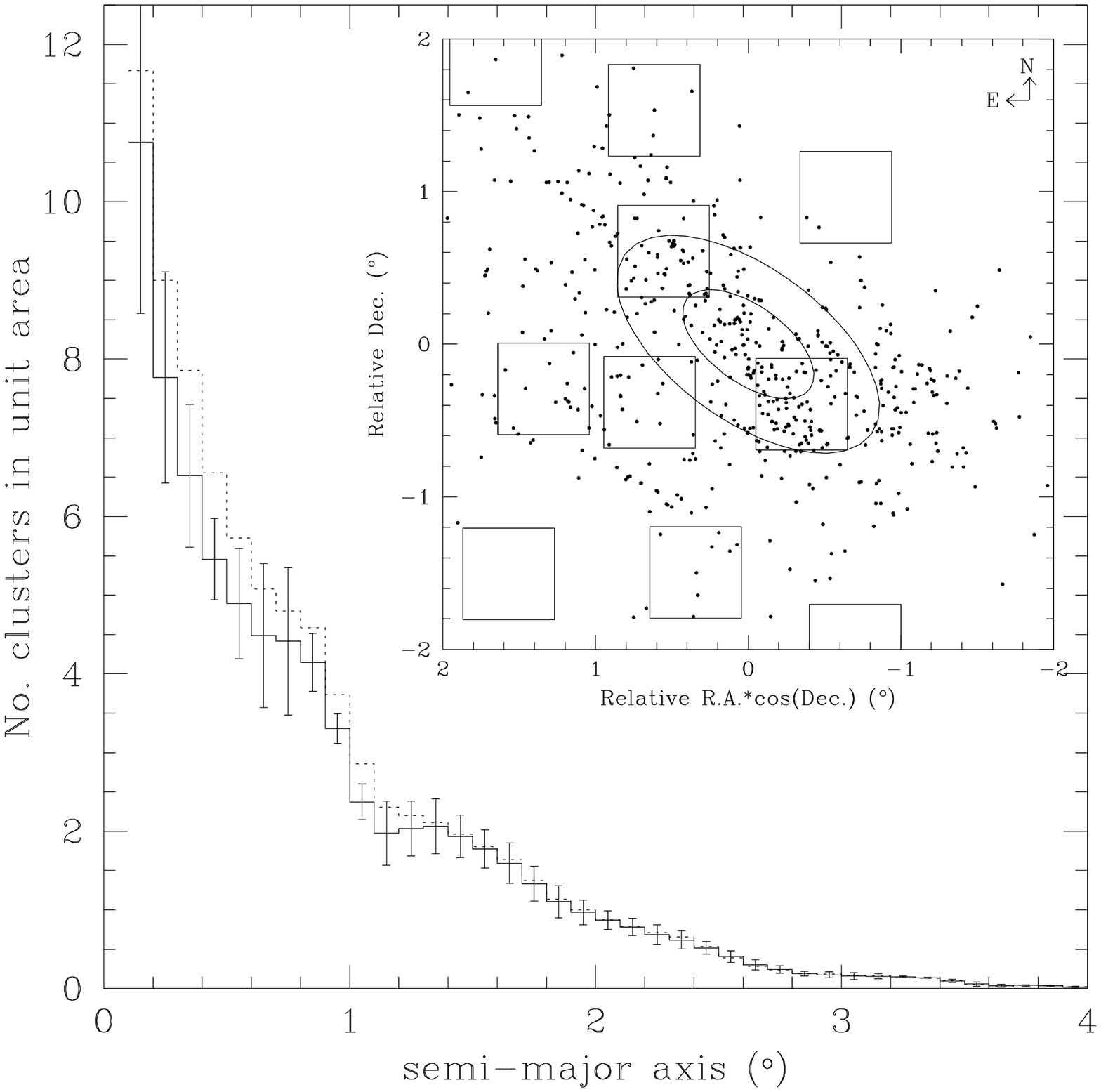,width=162mm}}
\caption{Spatial distribution of SMC clusters.}
\end{figure}

\end{document}